\documentclass[aps,prd,nofootinbib,preprint]{revtex4}
\usepackage[colorlinks=true, pdfstartview=FitV, linkcolor=blue, citecolor=red, urlcolor=magenta]{hyperref}
\usepackage{graphicx}
\usepackage{latexsym}
\usepackage{amsmath}
\usepackage{amsfonts}
\usepackage{amssymb}
\usepackage{verbatim}
\usepackage{dcolumn}
\usepackage{amssymb}
\usepackage{bm}
\usepackage{float}
\usepackage[english]{babel}
\newcommand{\be}{\begin{equation}}
\newcommand{\ee}{\end{equation}}
\newcommand{\bea}{\begin{eqnarray}}
\newcommand{\eea}{\end{eqnarray}}

\begin{document}

\newcommand{\JPess}{
\affiliation{Departamento de F\'isica, Universidade Federal da Para\'iba, \\Caixa Postal 5008, 58059-900, Jo\~ao Pessoa, PB, Brazil}
}

\newcommand{\Quixada}{\affiliation{
Universidade Estadual do Cear\'{a}, Faculdade de Educa\c{c}\~{a}o, \\Ci\^{e}ncias e Letras do Sert\~{a}o Central, 63900-000, Quixad\'{a}, CE, Brazil
}}

\title{Landau Levels in the Presence of a Cosmic String in Rainbow Gravity}

\author{V. B. Bezerra}
\email{valdir@fisica.ufpb.br}
\JPess
\author{I. P. Lobo}
\email{iarley\_lobo@fisica.ufpb.br}
\JPess
\author{H. F. Mota}
\email{hmota@fisica.ufpb.br}
\JPess
\author{C. R. Muniz}
\email{celio.muniz@uece.br}
\Quixada

%

\begin{abstract}
In this paper we analyze the energy levels of a charged scalar particle placed in the static cosmic string spacetime, under the action of a uniform magnetic field parallel to the string, in the context of the semi-classical approach of the rainbow gravity. Firstly, we focus on the non-relativistic regime by solving the corresponding Schr\"{o}dinger equation, following by a complete relativistic treatment of the problem in which we considered the Klein-Gordon equation. In both cases we find exact expressions for the Landau levels in terms of the rainbow functions, used to characterize a rainbow gravity model. In order to achieve the results of this paper we considered three different rainbow gravity models mostly used in the literature and compare the resulting modifications in the Landau levels with the standard case, namely without rainbow gravity. 

\end{abstract}


\maketitle
\section{Introduction}

Rainbow gravity is a semi-classical attempt of probing quantum gravity high-energy phenomena by introducing, by means of the so called rainbow functions, higher order terms in the energy-momentum dispersion relation, implying in a Lorentz symmetry breakdown at this energy scale \cite{Magueijo:2001cr,Magueijo:2002am,Magueijo:2002xx}. The rainbow functions depend on the ratio between the energy of a test particle (boson or fermion, for instance) and on the Planck energy, and govern departures from usual relativistic, minkowskian expressions. In this sense, in order to build this framework, the spacetime metric has also to be energy-dependent in such a way that this dependence becomes stronger as the energy of the probe particle approaches the Planck scale.

The rainbow gravity effective theory of quantum gravity, as it is also considered, is in fact an attempt to realize the curved spacetime generalization of Doubly Special Relativity (DSR) \cite{Magueijo:2001cr, Magueijo:2002am,AmelinoCamelia:2010pd,AmelinoCamelia:2000mn}. In the realm of the latter, modified dispersion relations are followed by deformed Lorentz symmetries in order to yield an invariant energy (length) scale, besides the usual invariant velocity of light (at low energies). Therefore, rainbow gravity also operates assuming as a fundamental principle that these two quantities are invariant, with the invariant energy scale set as the Planck energy \cite{AmelinoCamelia:2000mn, Magueijo:2001cr}.

From the mathematical structure point of view, DSR as formulated in \cite{Magueijo:2001cr,Magueijo:2002am} (which differs from other realizations of the DSR program, see for instance \cite{AmelinoCamelia:2000mn}) and consequently rainbow gravity, is accomplished by considering a non-linear transformation in momentum space, which induces a deformed Lorentz symmetry group that leads to corrections to the standard energy-momentum dispersion relation at the Planck scale. As it is known, the Planck length determines a threshold below which a continuous description of the spacetime loses its consistency, since quantum effects at such a scale are inevitably non-negligible and a classical spacetime metric description becomes impracticable.\footnote{In units such that $c=\hbar=1$, we have $E_{\text{Planck}}=\ell_{\text{Planck}}^{-1}$, therefore the length scale below which the spacetime becomes non-trivial can be translated as an energy threshold upon which high energy particles probes such quantum spacetime \cite{AmelinoCamelia:2008qg}.} In this context, a generalized uncertainty principle (GUP) also is often introduced in order to account for the fuzzy microscopic structure of the spacetime and to avoid the singularities of general relativity \cite{Maggiore:1993rv,Maggiore:1993zu}.

Rainbow gravity has received a great of attention in the past few years from both high-energies and gravitation communities. As such semi-classical approach is just an effective theory of quantum gravity, it is certainly incomplete, sharing with alternative proposals some common aspects, such as the search for a possible violation of the local Lorentz symmetry at very high-energy scales. In fact, departures from the standard Lorentz symmetry by modified dispersion relations can be found from less ad-hoc theoretical approaches, for instance from non-commutative geometry and loop quantum gravity \cite{Lukierski:1991pn,Majid:1994cy,Gambini:1998it}. Observationally, the motivation for the rainbow gravity approach is based on ultra-high energy cosmic rays \cite{Magueijo:2002am}, the phenomenology of TeV photons from Gamma Ray Bursts (GRBs) \cite{AmelinoCamelia:1997gz,grb1,Zhang:2014wpb} and recently from Ice Cube neutrinos \cite{Amelino-Camelia:2016ohi}, suggesting that the relativistic dispersion relation should be modified.

Our main objective here is to consider the Schr\"odinger and Klein-Gordon equations in the spacetime of a rainbow gravity cosmic string, that is, a cosmic string spacetime metric that depends on the energy of the probe particle through the rainbow functions.  In addition to that we also wish to consider a gauge field that provides an uniform magnetic field in the direction of the cosmic string. In other words, we want to investigate modifications in the Landau levels (energy spectrum) arising from this deformed cosmic string spacetime. The properties of a rainbow gravity cosmic string has been recently investigated by the authors in Ref. \cite{Momeni:2017cvl}.

Cosmic string is a linear topological defect expected to be formed as a consequence of phase transitions undergone by the universe in its very early stages \cite{escidoc:153364, hindmarsh,VS}. It is believed that cosmic strings may have important cosmological, astrophysical and gravitational implications making it a fruitful field of investigation \cite{Hindmarsh:2011qj,Mota:2014uka}. An additional motivation to consider cosmic string is the possibility to investigate some of its properties in laboratory through the condensed matter counterpart known as disclination and that is mostly found liquid crystals condensed matter systems \cite{2005cond.mat..2123K, Kleman:2008zz, Katanaev:1992kh}. Studies of Landau levels in the spacetime of a cosmic string (disclination) has been conducted in Refs \cite{Medeiros:2011zc, Furtado:1994np, Furtado:1999vd, DeAMarques:2001xbm, Cunha:2016uch}.\footnote{The effect of rainbow functions on energy levels has also been analyzed in the context of the Casimir effect in the Einstein universe \cite{Bezerra:2017zqq} and for the Dirac oscillator in \cite{Bakke:2018add}.}

The Landau levels have previously been considered in the context of modified theories of gravity, basically those ones involving the Generalized Uncertainty Principle, and can be found in Refs. \cite{Das:2008kaa,Ali:2011fa,Masood:2016wma}. The approach of these works is based on perturbation methods in quantum field theory. In this paper, we will choose another approach which seeks to solve exactly the Schr\"{o}dinger and Klein-Gordon equations in the cosmic string spacetime deformed by the rainbow gravity, extracting the energy levels from a simple quantization rule. Since cosmic strings are likely to be generated in the early universe, which is a period that might still present properties of a non-trivial spacetime, we aim to analyze these remnant traces in the form of rainbow functions in Landau energy levels.

This paper is organized as follows: in section II we discuss the main features of the rainbow gravity, in section III we compute the wave functions and the Landau levels of the scalar particle in both non-relativistic and relativistic regimes and, finally, in section IV we draw the conclusions.

\section{Main features of Rainbow Gravity}
\label{sec2}
As it has been said, the rainbow gravity is a generalization to curved spacetime of the deformed Lorentz symmetry group (locally), firstly studied in the context of DSR, and has as a consequence a modified energy-momentum dispersion relation in such way that high-order corrections can be investigated. Generally, the modification in the dispersion relation is presented in the form \cite{Magueijo:2001cr,Magueijo:2002am, Magueijo:2002xx}
\begin{equation}\label{MDR}
E^2f^2(\epsilon)-p^2c^2g^2(\epsilon)=m^2c^4,
\end{equation}
where $\epsilon=E/E_P$, with $E$ being the energy of the probe particle and $E_P$, the Planck energy. Thus, at high-energy scales, the functions $f(\epsilon)$ and $g(\epsilon)$ end up violating the standard energy-momentum dispersion relation, $g^{\mu\nu} p_{\mu}p_{\nu} = m^2c^2$, since the rainbow functions originates terms with order greater than $E^2$ in this relation. To describe a modified dispersion relation as a norm of the four-momentum in spacetime, the rainbow functions are absorbed in the vielbeins, leading an effective energy-dependent metric $g^{\mu\nu}(\epsilon)$. 
\par
On the other hand, it is evident that in the infrared regime
\begin{equation}
\lim_{\epsilon\rightarrow 0} f(\epsilon) = g(\epsilon)=1,
\label{inf}
\end{equation}
and therefore, rainbow gravity must recover standard general relativity in the infrared limit.
\par
Nowadays there are alternative formulations for a energy-dependent metric in the Planck scale in many contexts, for instance \cite{Barcaroli:2015xda,Pfeifer:2018pty,Carvalho:2015omv,Lobo:2016xzq,Lobo:2016lxm}. Along with rainbow gravity, these approaches present positive and negative features for a physically relevant metric description. Although rainbow gravity, as described here, turns out to be the most successful formalism due to its capacity to predict new effects in a large variety of deformed spacetimes.
\par
In the present work, we want to consider the three mostly adopted rainbow functions. The first of them is given by
\begin{equation}
f(\epsilon)=1,\qquad\qquad g(\epsilon) =\sqrt{1-\xi\epsilon^s}.
\label{rainf2}
\end{equation}
This rainbow function has also been considered, for instance, in Refs.\cite{Awad:2013nxa,Ali:2014xqa} to investigate the effects of the rainbow gravity on the Friedmann-Robertson-Walker (FRW) universe and for the regularization of the black hole evaporation due to Hawking radiation. Note that $\xi$ is an order one, dimensionless, free parameter of the formalism.

The second rainbow function we would like to consider is written as
\begin{equation}
f(\epsilon) = g(\epsilon) =  \frac{1}{1- \xi\epsilon}.
\label{rainf1}
\end{equation}
This rainbow function has been considered in Ref.\cite{Magueijo:2001cr,Magueijo:2002am} (see also references therein) and provides a constant velocity of light, and according to \cite{Magueijo:2002xx} may solve the horizon problem. It was also investigated the effects of the rainbow gravity on the FRW universe, in special, possible nonsingular universe solutions \cite{Awad:2013nxa}.

Finally, the third rainbow function to be considered is given by
\begin{equation}
f(\epsilon)=\frac{e^{\xi\epsilon - 1}}{\xi\epsilon},\qquad\qquad g(\epsilon) =1.
\label{rainf3}
\end{equation}
This rainbow function was considered in Ref.\cite{Awad:2013nxa,Santos:2015sva}, being originally proposed in \cite{AmelinoCamelia:1997gz} in one of the first papers regarding the use of GRBs to probe quantum gravity effects.

However, we should mention that it is also possible to derive observational consequences for general rainbow functions by considering some specific energies that should be reflected in numerical values assigned the functions, without necessarily relying on an specific form for them, see for instance the recent references in the context of black holes \cite{Hendi:2016hbe,Upadhyay:2018vfu}.

By adopting the three rainbow functions written above we want to obtain expressions for the infrared modification in the energy levels of a charged particle placed in the cosmic string spacetime deformed by the rainbow gravity, which is under the influence of an uniform magnetic field parallel to the string. 

\section{Landau Levels in the rainbow gravity Cosmic String Spacetime}

The study of the Landau levels in the presence of a standard cosmic string (or a disclination in condensed matter systems) was already performed in \cite{Medeiros:2011zc, Furtado:1994np, Furtado:1999vd, DeAMarques:2001xbm, Cunha:2016uch}, and we will follow the same line of investigation. In this case, the Hamiltonian corresponding to a charged particle in the presence of a gauge field in a curved background given generically by a metric $g_{ij}$ is written as
\begin{equation}
\hat{H}=\frac{1}{2m \sqrt g}(-i\hbar\partial_i-\frac{e}{c} A_i)[\sqrt g g^{ij}(-i\hbar\partial_j-\frac{e}{c}
A_j)]\ , \label{f11}
\end{equation}
where $g=\text{det} (g_{ij})$ stands for the determinant of the metric $g_{ij}$. Note that we have considered, in the case in which exists a minimal coupled gauge field $A_i$, a generalization of the Laplace operator in curved space given by
\begin{eqnarray}
\nabla^{2} = \frac{1}{\sqrt{g}} \partial_{i} \left( g^{ij} \sqrt{g}
\partial_{j}\right). \hspace{0.5cm}
\label{LO}
\end{eqnarray}
The gauge field is minimally coupled (in the electromagnetic sense) to the momentum operator in the following way: $ -i\hbar\partial_i\mapsto -i\hbar\partial_i -
qA_i$, with $e$ being the electric charge of the particle.

As it has been reported in \cite{Momeni:2017cvl}, the spacetime of a cosmic string modified by the rainbow gravity is given by
\begin{equation} \label{CosmicStringRainbowMetric}
ds^{2} =\frac{c^2dt^2}{f^2(\epsilon)} - \frac{1}{g^2(\epsilon)}\left(dz^{2} + d \rho^{2} + \alpha^{2}
\rho^{2} d \phi^{2}\right).
\end{equation}
In fact, this line element describes a spacetime which is one of the non-Kasner family of exact solutions found in \cite{Momeni:2017cvl}, corresponding to the vacuum of the Levi-Civita spacetime. One of several aspects of the debate that initially arose about rainbow gravity was whether it is possible to re-scaling the time as $t\rightarrow t/f(\epsilon)$ and the spatial coordinates $\rho$ and $z$ as $\rho\rightarrow \rho/g(\epsilon)$, $z\rightarrow z/g(\epsilon)$, considering that the particle energy is a constant. In this way, for instance, the spacetime described by the line element \eqref{CosmicStringRainbowMetric} would be essentially the line element describing the standard cosmic string spacetime. However, we can note that the energy dependence in rainbow gravity spacetimes leads to a reconsideration of the measurement process (which is physical) and, therefore, one cannot say that this is just a matter of mere mathematical re-scaling of both time and spatial coordinates. As it has been known, the physics arising from the rainbow gravity framework deeply affects several aspects of the classical concepts of general relativity, with the metric dependence on the particle energy being seen as a kind of backreaction effect, as expected from a genuine quantum gravity theory.
\subsection{Schr\"{o}dinger equation}
Let us now consider the Schr\"{o}dinger equation in order to determine the Landau levels. Although the Schr\"{o}dinger equation describes the non-relativistic aspect of a charged particle, we should be able to get relativistic corrections to the Landau levels due to fact that we are considering a semi-classical quantum gravity approach, which is itself relativistic. It is worth to recall here that the rainbow gravity modifications occur at the metric level.  Thus, the Schr\"{o}dinger equation in a curved space can then be written by making use of the Hamiltonian operator \eqref{f11}, that is,
\begin{equation}
\label{se}
\hat{H}\Psi=i\hbar\frac{\partial\Psi}{\partial t}.
\end{equation}
The Schr\"odinger equation written in this way takes into account the curved spatial part of the line element \eqref{CosmicStringRainbowMetric} and incorporates the non-trivial topology of the cosmic string, as well as the modifications brought by the rainbow gravity framework.

In order to solve the Schr\"odinger equation \eqref{se} and find the corresponding modified Landau levels we will consider a configuration of the gauge field, $A_i$, such that its only nonzero component provides an uniform magnetic field in the $z$-direction, along which the cosmic string is placed, i.e.,
\begin{equation}
A_{\phi}(\rho)=\frac{B\rho}{2\alpha}\ . \label{f10}
\end{equation}
Thus, the Hamiltonian now becomes

\begin{eqnarray}
\hat{H}&=&-\frac{\hbar^2g^2(\epsilon)}{2m}\left\{\frac{\partial^2}{\partial
z^2}+\frac{1}{\rho}\frac{\partial}{\partial\rho}\left(\rho\frac{\partial}{\partial\rho}\right)+
\frac{1}{\alpha^2\rho^2}\frac{\partial^2}{\partial\phi^2}\right\}\nonumber\\
&+&\frac{i\hbar\,q\,B g(\epsilon)}{2\alpha^2m}\frac{\partial}{\partial\phi}+
\frac{q^2B^2\rho^2}{8m\, \alpha^2}.  \label{f11}
\end{eqnarray}
%
We can see that, in the limits $\alpha \rightarrow 1$ and $f(\epsilon)$, $g(\epsilon)\rightarrow 1$, the expression (\ref{f11}) is the
Euclidean Hamiltonian in the presence of a coupled gauge field given by \eqref{f10}. The solution of the time independent Schr\"odinger equation
corresponding to the Hamiltonian (\ref{f11}) can be written in the stationary wave form $\Psi = e^{-i\frac{E}{\hbar}t}\psi$, where the solution $\psi(z,\rho,\phi)$ for the spatial part can be written as
\begin{equation}
\psi(\rho,\phi,z)=e^{ikz}e^{i\ell\phi}R(\rho)\ , \label{f12}
\end{equation}
\noindent
in accordance with the translational symmetry along the $z$-direction as well as the azimuthal one around the string. In other works, both the momenta in the $z$ and $\phi$ directions commute with the Hamiltonian \eqref{f11}. Thus, the radial equation is found to be
\begin{eqnarray}
-\frac{\hbar^2g^2(\epsilon)}{2m}\left\{-k^2R(\rho)+\frac{1}{\rho}\frac{d R(\rho)}{d\rho}+\frac{d^2 R(\rho)}{d\rho^2}-
\frac{\ell^2}{\alpha^2\rho^2}R(\rho)\right\}\nonumber\\
-\frac{\hbar\,q\,B g(\epsilon)\, \ell}{2\alpha^2m}R(\rho)+
\frac{q^2B^2\rho^2}{8m\, \alpha^2}R(\rho)-E\, R(\rho)=0,
\end{eqnarray}
whose solution is given in terms of the confluent hypergeometric function, ${}_1F_1(\beta;\gamma;x(\rho))$, is given by
\begin{eqnarray}
R(\rho)=C\exp\left(-\frac{|q|B\rho^2}{4g(\epsilon)\hbar\alpha}\right)
\rho^{\frac{|\ell|}{\alpha}}
{}_1F_1\left(\beta,\frac{|\ell|}{\alpha}+1,
\frac{|q|B\rho^2}{2\hbar\alpha g(\epsilon)}\right)\ , \label{f133}
\end{eqnarray}
where
\begin{equation}
\beta=\frac{\alpha^2 g^2(\epsilon)\hbar^2k^2+B\alpha g(\epsilon)\hbar q-2E\alpha^2  m+(\ell-|\ell|)Bg(\epsilon)\hbar q}{2\alpha g(\epsilon)\hbar Bq},
\end{equation}
and  $C$ is a normalization constant. Note that the radial solution (\ref{f133}) has a dependence on the rainbow function $g(\epsilon)$ through the arguments of both the exponential and confluent hypergeometric functions.

The energy (Landau) levels are obtained by the requirement that the radial solution must be regular at infinity, which implies that the confluent hypergeometric function turns to be a polynomial of degree $n$ in $x(\rho)$. This procedure is performed in our case by making $\beta=-n$, with $n=0,1,2,3$. This provides the rainbow function-dependent energy levels
\begin{equation}
\epsilon_{n\ell}=\frac{\hbar \omega_{B}g(\epsilon_{n\ell})}{2 \alpha E_P }
\left(2n+\frac{|\ell|}{\alpha}-\frac{\ell}{\alpha}+1\right)+
\frac{\hbar^2k^2g^2(\epsilon_{n\ell})}{2mE_P}, \label{f14}
\end{equation}

\noindent
with $\ell=0,\pm1,\pm2$..., and
$\omega_{B}=\frac{|q|B}{m }$ is the cyclotron frequency. We can clearly see in Eq. \eqref{f14} the modification of the Landau levels due to its dependence on the rainbow function $g(\epsilon)$. Note, however, that in the limits $\alpha \to 1$ and $f(\epsilon)$, $g(\epsilon)\to 1$, Eq. (\ref{f14}) gives the Landau levels plus the kinetic energy corresponding to the free motion along the $z$-axis in the Minkowski spacetime. These energy levels have already been considered previously by the authors in Refs.  \cite{Medeiros:2011zc, Furtado:1994np, Furtado:1999vd, DeAMarques:2001xbm, Cunha:2016uch}, which shows the consistency of our results.

We shall next consider the three models of rainbow gravity commonly adopted in the literature in order to try to explicitly compute the Landau levels of a charged particle in the cosmic string spacetime. As we are considering the non-relativistic (i.e. $\epsilon\ll 1$) dynamics of a charged particle, through the Schr\"odinger equation, the relativistic corrections will be obtained via the rainbow functions for each specific gravity model. 
\par
This way, the results of this section should rely on low values of $\epsilon$ and the cyclotron frequency $\omega_B$, however, in order to graphically demonstrate the qualitative influence of the rainbow functions on the energy levels, we will extrapolate the non-relativistic regime by considering trans-planckian values of $\omega_B$ for values of the parameter $\alpha$ of the order $1$ and $10^{-1}$. As expected, for smaller values of $\alpha$, which corresponds to a stronger gravitational interaction, quantum gravity effects become relevant for smaller values of $\omega_B$, as can be verified from Eq.(\ref{f14}), in which case the relevant variable is $\hbar\omega_B/E_P\alpha$. In other words, the stronger the gravitational field is, more evident becomes the effects of rainbow gravity.
\\

{\bf (1) Case:} $f(\epsilon)=1$ and $g(\epsilon)=\sqrt{1-\xi\epsilon^s}$
\\

\begin{figure}[H]
\centering
\includegraphics[scale=0.5]{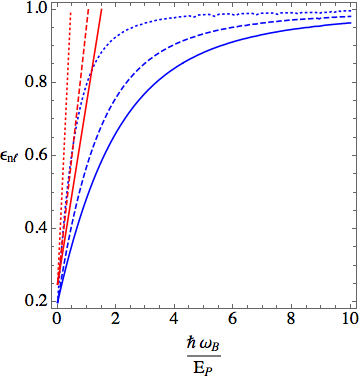}
\caption{Plot of the energy levels for $(n,\ell)=0$ as a function of $\hbar\omega_B/E_P$. The red curves correspond to the undeformed case $f=g=1$, while the blue ones we set $f=1$ and $g=\sqrt{1-\xi\epsilon^s}$. For both cases, solid lines represent the flat spacetime $\alpha=1$, while dashed and dotted ones represent $\alpha=0.7$ and $0.3$, respectively. We set $c=1$ and $\hbar^2k^2/mE_P=0.5$.}
\label{sch1}
\end{figure}

In this model, it is easy to see from Eq. (\ref{f14}) that the rainbow gravity will affect both the translational and rotational parts of the energy. In order to get relativistic corrections to the Landau levels let us consider $s=1$. In this case, we find the following first order (in $\xi\epsilon$) approximated expression for the Landau levels:
\begin{equation} \label{LandauRainbow1}
\epsilon_{n\ell}\approx \epsilon_{n\ell}^{(0)}\left[1-\frac{\xi}{2}\left(\epsilon_{n\ell}^{(0)}+\frac{\hbar^2k^2}{2mE_P}\right)\right],
\end{equation}
where $\epsilon_{n\ell}^{(0)}$ are the Landau levels in the absence of the rainbow gravity  \cite{Medeiros:2011zc, Furtado:1994np, Furtado:1999vd, DeAMarques:2001xbm, Cunha:2016uch}. One should point out that the effect of the rainbow gravity is to decrease the energy levels when compared to the standard case. This can be globally verified by Fig.(\ref{sch1}), where one can see the exact behavior of the energy levels as a function of the cyclotron frequency $\omega_B$ (in units of $E_P/\hbar$), where the deformed energy (in blue) for fixed quantum numbers $(n,\ell)$, is always smaller than the undeformed one (in red). Furthermore, it can be verified the upper bound in $\epsilon$, which is a feature of this rainbow function. The effect of considering different quantum numbers consists in a quantitative shift of these graphs, without any qualitative difference between them. Therefore, we set $(n,\ell)=0$ and we considered the cases $\alpha=1$ as the solid curves, which consists in the absence of the cosmic string; and $\alpha=0.7$ (dashed line) and $\alpha=0.3$ (dotted line). In these cases, one can see a decrease of the energies, as expected from Eq.(\ref{f14}).

{\bf (2) Case:} $f(\epsilon)=g(\epsilon)=(1-\xi\epsilon)^{-1}$
\\

\begin{figure}[H]
\centering
\includegraphics[scale=0.5]{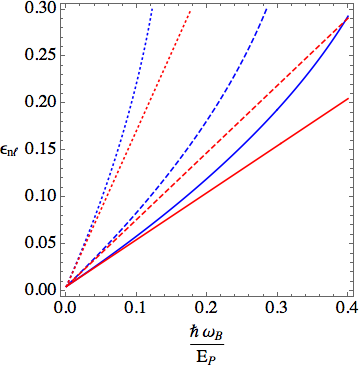}
\caption{Plot of the energy levels as a function of $\hbar\omega_B/E_P$. The red curves consist in the undeformed case $f=g=1$, and for the blue ones we set $f=g=(1-\xi\epsilon)^{-1}$. Solid lines represent the flat spacetime $\alpha=1$, while dashed and dotted ones represent $\alpha=0.7$ and $0.3$, respectively. We set $c=1$ and $\hbar^2k^2/mE_P=0.01$.}
\label{sch21}
\end{figure}

This case can also be expanded up to first order in $\xi\epsilon$, such that we find
\begin{equation} \label{LandauRainbow2}
\epsilon_{n\ell}\approx \epsilon_{n\ell}^{(0)}\left[1+\xi\left(\epsilon_{n\ell}^{(0)}+\frac{\hbar^2k^2}{2m E_P}\right)\right].
\end{equation}
Differently from the result in Eq.\eqref{LandauRainbow1}, the one in Eq.\eqref{LandauRainbow2} shows that the effect of the rainbow gravity correction is to increase the Landau levels. This behavior can be seen in Fig.(\ref{sch21}), where as before the straight red lines refer to the cases $f(\epsilon)=g(\epsilon)\equiv 1$, and the blue ones are the deformed energy levels. In this case, even though there is the same kind of bound in $\epsilon$ as before, the shapes of the curves are completely different and the effect of the rainbow functions consists in increase the energies for fixed quantum numbers. Obviously, this behavior can only be verified within the energy's limit. 
\par
In this case, we can see a degeneracy due to the cyclotron frequency, i.e., the same $\omega_B$ leads to two different energy levels. This feature will be removed once we consider the relativistic case of the Klein-Gordon equation in the next section.

\begin{figure}[H]
\centering
\includegraphics[scale=0.5]{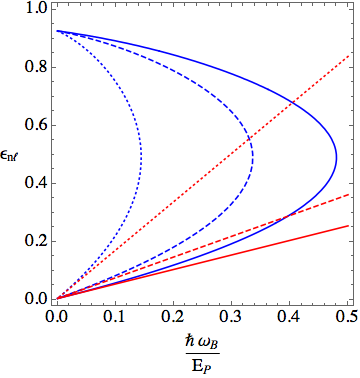}
\caption{The energy levels present a degeneracy in the cyclotron frequency $\omega_B$ at the planckian regime.}
\label{sch2}
\end{figure}

In fact, if we consider the planckian regime, which corresponds to $\epsilon_{n\ell}$ of order one, we shall verify a degeneracy in $\omega_B$ for the energy levels since Eq.\eqref{f14} would not be a function anymore, as can be verified in Fig.(\ref{sch2}). As we will see, such undesirable feature is only present in this case of the Schr\"odinger equation, and the Klein-Gordon case, that we will consider in the next section, this feature will be absent of this property.

The third case to be considered would be with the rainbow functions given by Eq. (\ref{rainf3}). However, here we obtain the trivial result $\epsilon_{n\ell}=\epsilon_{n\ell}^{(0)}$, i.e., the Landau levels are not modified by the rainbow gravity.
%
\subsection{Klein-Gordon equation}
%
Now we want to analyze the effects of the rainbow gravity models introduced in Sec.\eqref{sec2}, in the relativistic Landau levels by solving the Klein-Gordon equation in the background spacetime under consideration \cite{Momeni:2017cvl}. In order to do that, let us then consider the dispersion relation $E^2-p^2c^2=m^2c^4$ with the electromagnetic minimal coupling $p_{\mu}\mapsto p_{\mu}-qA_{\mu}$. The resulting Klein-Gordon equation in a curved spacetime is obtained by considering the standard quantization procedure $(E,p_i)\mapsto (i\hbar\, \partial_t,-i\hbar\, \partial_i)$, with the Laplace operator given by Eq. \eqref{LO}, which results in,
\begin{equation}
-\frac{\hbar^2f^2}{c^2}\frac{\partial^2\Psi}{\partial t^2}+\hbar^2g^2\left\{\frac{\partial^2}{\partial
z^2}+\frac{1}{\rho}\frac{\partial}{\partial\rho}\left(\rho\frac{\partial}{\partial\rho}\right)+
\frac{1}{\alpha^2\rho^2}\frac{\partial^2}{\partial\phi^2}\right\}\Psi-\frac{i\hbar\,q\,B g}{\alpha^2}\frac{\partial\Psi}{\partial\phi}-
\frac{q^2B^2\rho^2}{4\, \alpha^2}\Psi=m^2c^2\Psi.
\end{equation}
where the gauge field \eqref{f10} has been used. Again, we can consider the stationary wave solution form $\Psi = e^{-i\frac{E}{\hbar}t}\psi$, with $\psi(\rho, \phi,z)$ given by \eqref{f12}. Thus, we have a second order differential equation for $R(\rho)$:
\begin{eqnarray}
f^2\frac{E^2}{c^2}R(\rho)+\hbar^2g^2\left(-k^2R(\rho)+\frac{1}{\rho}\frac{dR(\rho)}{d\rho}+\frac{d^2R(\rho)}{d\rho^2}-\frac{\ell^2}{\alpha^2\rho^2}R(\rho)\right)\nonumber\\
+\frac{\hbar\,q\,B\, g\, \ell}{\alpha^2 }R(\rho)-\frac{q^2B^2\rho^2}{4\, \alpha^2}R(\rho)-m^2c^2R(\rho)=0.
\end{eqnarray}
The solution is also given in terms of the confluent hypergeometric function, as follows,
\begin{eqnarray}
R(\rho)=C\exp\left(-\frac{|q|B\rho^2}{4g(\epsilon)\hbar\alpha}\right)
\rho^{\frac{|\ell|}{\alpha}}
{}_1F_1\left(\beta,\frac{|\ell|}{\alpha}+1,
\frac{|q|B\rho^2}{2\hbar\alpha g(\epsilon)}\right)\ , \label{f13}
\end{eqnarray}
where
\begin{equation}
\beta=\frac{\alpha^2 c^2\hbar^2k^2g^2(\epsilon)+\alpha^2 c^4m^2+\alpha qBc\hbar g(\epsilon) -E^2\alpha^2 f^2(\epsilon)+(|\ell|-\ell)qBc\hbar g(\epsilon)}{2\alpha c^2 \hbar Bq g(\epsilon)},
\end{equation}
and  $C$ is a normalization constant. Note that the parameter $\beta$ also depends on the rainbow function $f(\epsilon)$ in this case. As before, by
imposing the regularity condition $\beta=-n$ for the confluent hypergeometric function, we find that the energy spectrum $\epsilon_{n\ell}=E_{n\ell}/E_P$ is given by
\begin{equation}
\epsilon_{n\ell}=\frac{1}{\alpha E_Pf(\epsilon)}\sqrt{\alpha^2 c^4m^2+g(\epsilon)[\alpha^2 c^2\hbar^2k^2g(\epsilon)+2n\alpha mc^2\hbar\omega_B +(|\ell|-\ell)|q|Bc\hbar +\alpha mc^2\hbar\omega_B]}.
\label{RLL}
\end{equation}
The energy levels above are the generalization of the Landau levels in the standard cosmic string spacetime \cite{Medeiros:2011zc, Furtado:1994np, Furtado:1999vd, DeAMarques:2001xbm, Cunha:2016uch} to the case where the cosmic string is modified by the rainbow gravity according to the line element \eqref{CosmicStringRainbowMetric}. Thereby, we can see that the Landau levels above, in fact, depend on the rainbow functions $f(\epsilon)$ and $g(\epsilon)$. To be more precise, let us next analyze the Landau levels \eqref{RLL} for the three rainbow functions presented in Sec.\eqref{sec2}.
\\

{\bf (1) Case:} $f(\epsilon)=1$ and $g(\epsilon)=\sqrt{1-\xi\epsilon^s}$
\\

\begin{figure}[H]
\centering
\includegraphics[scale=0.5]{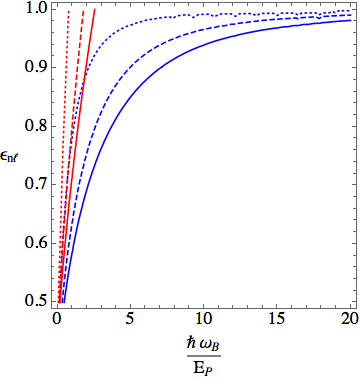}
\caption{Plot of the energy levels for $(n,\ell)=0$ as a function of $\hbar\omega_B/E_P$. The red curves consist in the undeformed case $f=g=1$, while for the blue ones we set $f=1$ and $g=\sqrt{1-\xi\epsilon^s}$. For both cases, solid lines represent the flat spacetime $\alpha=1$, while dashed and dotted ones represent $\alpha=0.7$ and $0.3$, respectively. We set $c=1$ and $m^2/E_P^2=\hbar^2k^2/E_P^2=0.1$.}
\label{klein1}
\end{figure}

In this case we consider $s=1$ and take an approximation in $\xi$ of \eqref{RLL} up to first order. This provides
\begin{equation}
\epsilon_{n\ell}\approx \epsilon^{(0)}_{n\ell}-\frac{\xi}{4}\left[\left(\epsilon^{(0)}_{n\ell}\right)^2-\frac{m^2c^4}{E^2_P}+\frac{c^2\hbar^2k^2}{E^2_P}\right]<\epsilon^{(0)}_{n\ell}.
\end{equation}
At first order in $\xi$, the Landau levels is smaller than in the case without rainbow gravity when the Landau levels is given by \eqref{RLL} for $f(\epsilon)=g(\epsilon)=1$, that is, $\epsilon^{(0)}_{n\ell}$. Again, this is confirmed by the global behavior of the energy levels in Fig.(\ref{klein1}), which is qualitatively similar to the the Schr\"odinger case. Again, the energy bound can explicitly seen in this graph, and the choice of non-null quantum numbers simply translates the curves without modifying their shapes.

{\bf (2) Case:} $f(\epsilon)=g(\epsilon)=(1-\xi\epsilon)^{-1}$
\\

\begin{figure}[H]
\centering
\includegraphics[scale=0.5]{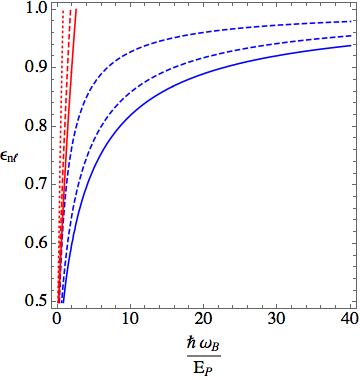}
\caption{Energy levels as a function of $\hbar\omega_B/E_P$. The red curves consist in the undeformed case $f=g=1$, and the blue ones we set $f=g=(1-\xi\epsilon)^{-1}$. For both cases, solid lines represent the flat spacetime $\alpha=1$, while dashed and dotted ones represent $\alpha=0.7$ and $0.3$, respectively. We set $c=1$ and $m^2/E_P^2=\hbar^2k^2/E_P^2=0.1$.}
\label{klein2}
\end{figure}

In this case, we also get an approximation, up to first order, to the Landau levels \eqref{RLL}:
\begin{equation}
\epsilon_{n\ell}\approx \epsilon^{(0)}_{n\ell}-\frac{\xi}{2}\left[\left(\epsilon^{(0)}_{n\ell}\right)^2+\frac{m^2c^4}{E^2_P}-\frac{c^2\hbar^2k^2}{E^2_P}\right]<\epsilon^{(0)}_{n\ell}.
\end{equation}
Again, the Landau levels is smaller then in the standard case, without rainbow gravity. Here, the degeneracy on the cyclotron frequency presented in the Schr\"odinger case is removed when we consider the Klein-Gordon equation. The qualitative behavior of this case is the same as the previous one, however the curves reach the asymptote line in a slower rate in this case as can be seen in Fig.(\ref{klein2}).
Here, the deformed eigenenergies are smaller than the undeformed ones, which is a behavior opposite to the Schr\"odinger case. The reason is the presence of the ``rest mass" contribution in this case, and this will be valid for any context. In fact, consider a free particle with the modified dispersion relation of these rainbow functions
\be
E^2-c^2p^2=(1-\epsilon)^2m^2c^4,
\ee
which is equivalent as if we have an energy-dependent mass $m(\epsilon)=(1-\epsilon)m$. In this case, the non-relativistic regime of this dispersion relation reads
\be\label{appr-energy}
E\approx m(\epsilon)c^2+\frac{p^2}{2m(\epsilon)}.
\ee

Due to the energy bound, the undeformed mass (which we call $m_0$) is always larger than the deformed one, $m(\epsilon)\leq m_0$. Therefore the non-relativistic kinetic energy satisfies the inequality
\be
\frac{p^2}{2m(\epsilon)}\geq \frac{p^2}{2m_0},
\ee
which corresponds, qualitatively, to the result obtained in the previous section, that the rainbow-corrected eigenenergy is larger that the undeformed one in the non-relativistic regime. This qualitative analysis was done for the free particle, but as we saw it is valid if we consider the presence of gravitational and electromagnetic fields.
\par
For the Klein-Gordon equation, we are also considering contributions due to the ``rest mass", whose dominant contribution to the energy imposes the opposite inequality that we observed in this example.
\\
\\
{\bf (3) Case:} $f(\epsilon)=\frac{(e^{\xi\epsilon}-1)}{\xi\epsilon}$ and $g(\epsilon)=1$
\\

\begin{figure}[H]
\centering
\includegraphics[scale=0.5]{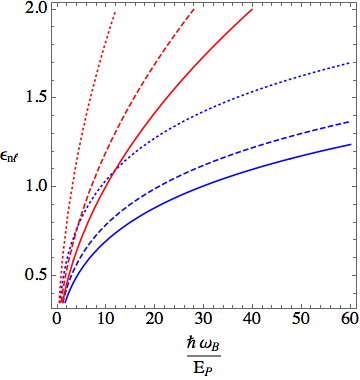}
\caption{Energy levels for $(n,\ell)=0$. The red curves consist in the undeformed case $f=g=1$, and for the blue ones we set $f=\frac{(e^{\xi\epsilon}-1)}{\xi\epsilon}$ and $g=1$. As before, solid lines represent the flat spacetime $\alpha=1$, while dashed and dotted ones represent $\alpha=0.7$ and $0.3$, respectively. We set $c=1$ and $m^2/E_P^2=\hbar^2k^2/E_P^2=0.01$.}
\label{klein4}
\end{figure}

The approximation for this case takes the form
\begin{equation}
\epsilon_{n\ell}\approx \epsilon^{(0)}_{n\ell}-\frac{\xi}{2}\left(\epsilon^{(0)}_{n\ell}\right)^2<\epsilon^{(0)}_{n\ell}.
\end{equation}
We can see one more time that the modified Landau levels are smaller than in the standard case. Since there is no energy-bound for these rainbow functions, they can grow indefinitely accompanying the increase of the frequency. However, in this case, this growth is strongly attenuated in comparison to the undeformed metric, i.e., it would require high values of the frequency in order to substantially increase the energy, as can be seen in Fig.(\ref{klein4}).

Therefore, in all three relativistic cases, the Landau levels are being pulled down by the rainbow functions, which was not totally the case of the non-relativistic analysis. This behavior is in accordance with previous analysis of the effect of rainbow function on energy levels in other contexts \cite{Bezerra:2017zqq}.

\section{Concluding Remarks}

We have computed the wave functions and the energy levels of a charged scalar particle placed in the static cosmic string spacetime, which is under the action of an uniform magnetic field parallel to the string, in the scenario of the rainbow gravity. Firstly, we have focused on the non-relativistic regime by solving the corresponding Schr\"{o}dinger equation. In this case, it was found a relativistic correction for the particle energy in the first order of the rainbow gravity free parameter $\xi$, which comes from the deformation of the standard cosmic string metric due to the rainbow functions. In this approximation, we have shown that when one considers the functions given by Eq. (\ref{rainf2}), the corresponding Landau levels are lower than those ones obtained in the standard case, without the rainbow gravity. On the other hand, by taking into account the rainbow functions of Eq. (\ref{rainf1}), these energies are greater than the ones of the standard case based on general relativity. The functions given by Eq. (\ref{rainf3}) showed no changes in the Landau levels when compared with this case.

The complete relativistic treatment of the problem was also done by solving the Klein-Gordon equation and by considering the same first order correction in $\xi$ parameter. Unlike the previous situation we have found that all the rainbow models analyzed here lowered the eigenvalues of energy with respect to the standard case. Thus, from the point of view of the energy of the system and in the regime under consideration, the nature favors the rainbow gravity in detriment of the general relativity.

Another remarkable feature that we can notice from the models studied in both the regimes is that the influence of the rainbow gravity on the Landau levels increases with the intensity of the magnetic field and with the conicity of the spacetime, i.e., when $\alpha$ approaches to zero. As a future perspective to this work, we intend to study that problem by considering charged fermions.

\acknowledgments
The authors would like to thank CNPq (Conselho Nacional de Desenvolvimento Cient\'ifico e Tecnol\'ogico - Brazil) for financial support. This study was financed in part by the Coordena\c{c}\~ao de Aperfei\c{c}oamento de Pessoal de N\'ivel Superior - Brasil (CAPES) - Finance Code 001.

\end{document}